\documentclass[conference]{IEEEtran}
\IEEEoverridecommandlockouts

\usepackage{cite}
\usepackage{amsmath,amssymb,amsfonts}
\usepackage{algorithmic}
\usepackage{graphicx}
\usepackage{textcomp}
\usepackage{xcolor}
\usepackage{fancyhdr}
\usepackage{lipsum} 
\usepackage{mathtools}
\usepackage{hyperref}
\usepackage{longtable}  

\usepackage{pdflscape} 

\def\BibTeX{{\rm B\kern-.05em{\sc i\kern-.025em b}\kern-.08em
    T\kern-.1667em\lower.7ex\hbox{E}\kern-.125emX}}
    
\fancypagestyle{firstpagefooter}{%
  \fancyhf{}
  
  \fancyfoot[R]{Menelaos-NT Research Report template by Zhouyan Qiu, University of Vigo}
}

\begin{document}
\title{EIE calculation and Collisional-Radiative modeling for Na-like Kr and Xe}

\author{\IEEEauthorblockN{Ashwini Malviya}
\IEEEauthorblockA{\textit{Department of Physics} \\
\textit{IIT Roorkee}\\
Roorkee, India \\
a\_malviya@ph.iitr.ac.in}

}
\twocolumn[
\begin{@twocolumnfalse}
\maketitle
\end{@twocolumnfalse}

\begin{abstract}
As an extension to our previous work \cite{article}, a comprehensive theoretical study for Na-like Krypton and Xenon is carried out. 
Using MCDHF(Multiconfiguration Dirac-Hartee-Fock) along with RDW(Relativistic distorted wave) theory we calculate key atomic properties, electron-impact excitation (EIE), rate coefficients, and collision strength for these ions. We use these parameters to build a Collisional-Radiative model for Na-like Krypton and Xenon. For Na-like Krypton we compare our computed excitation energy, EIE cross-sections,  rate coefficients, emission line intensity with previous work. Additionally we investigate variation of line ratios with temperature. For Na-like Xenon we compared excitation energy for various fine-structure with NIST(\cite{nist_atomic_spectra}) database and then provide our computed results for EIE cross-section,intensity profile, and the temperature dependence of line ratios for Na-like Xenon. Our findings offer atomic data for studies related Na-like ions.     
~
\end{abstract}

]
\section{Introduction}
Argon, Krypton, and Xenon are important elements in atomic physics, astrophysics, and fusion research due to the unique properties of their highly charged ions. These ions are especially relevant in magnetic fusion contexts, such as in tokamak reactors, where they can serve as diagnostic tools through the analysis of spectral lines emitted by high-temperature plasma \cite{Beiersdorfer_2015}. Accurate atomic structure parameters for these ions are critical for interpreting their spectral properties, which in turn improves our understanding of plasma behavior in experimental settings \cite{Hill2022SummaryRO}. In astrophysics, the discovery of argon in hot, dense stars highlights the importance of these ions for studying stellar phenomena \cite{refId0}.

Considering the significance of these ions, substantial efforts have been made to calculate accurate atomic parameters \cite{PhysRevA.102.062823,PhysRevA.55.3229,Liang_2009,osti_69419,refId0,E_Charro_2002,S_Bliman_1989,N_Reistad_1986,Osin_2012,Rice_2020,article, kr}. To advance the database of atomic parameters for these ions and improve predictive models, 
building on our previous research of calculating atomic parameters related to Na-like ions, this work extends our earlier efforts by performing detailed calculations of EIE cross-sections for Na-like Kr and Xe and constructing a comprehensive CR model for Na-like Xe, giving insight into intensity spectra and variations of line with temperature. To validate our calculation, we  compared our results with previous work \cite{kr} for Na-like Kr.
The calculations are performed using the Multiconfiguration Dirac–Hartree–Fock (MCDHF) and Relativistic Distorted Wave (RDW) theories, which are briefly outlined in the following section along with the details of CR model.  


\section{Method}
\subsection{MCDHF theory}
In multiconfiguration Dirac-Hartee-Fock theory \cite{grant2007relativistic}, we formulate the atomic state function (ASF) as a linear combination of Configuration state functions (CSFs); CSFs are given by an antisymmetric product of single electron Dirac orbitals.
\begin{equation}
\Psi(\gamma PJM) = \sum_{j=1}^{n_C} c_j \Phi(\gamma_j PJM)
\end{equation}
where $\psi$ is ASF, \textit{P, J, M} are parity, total angular momentum, and its projection. $\phi$ is CSF, and $c_j$ is the mixing coefficient, representing the corresponding CSFs' contribution.  
The single electron orbitals are given by Dirac-Hartee-Fock equations; these equations incorporate the interaction of electrons with a nucleus and the mean field of another electron. The Dirac-Coulomb Hamiltonian involved is given by 
\begin{equation}
\hat{H}_{DC} = \sum_{i=1}^{N} \left[ c (\boldsymbol{\alpha} \cdot \mathbf{p}_i) + (\beta - I)c^2 + V_i \right] + \sum_{i<j} \frac{1}{r_{ij}}
\end{equation}
where c, is speed of light, $\alpha, \beta$ are Dirac matrices, $p_i$ momentum operator,$V_i$ represent potential energy, and $\frac{1}{r_{ij}}$ is coulumb repulsion. MCDHF uses variational principle to minimize the total energy given by 
\begin{equation}
E[\Psi] = \frac{\langle \Psi | \hat{H}_{DC} | \Psi \rangle}{\langle \Psi | \Psi \rangle}
\end{equation}
this optimization is performed by adjusting the mixing coefficient and orbitals. 
The transition matrix element of the radiative operator T provides the transition probability; the rate is given by 
\begin{equation}
A_{i \to f} = \frac{2\pi}{2J_i + 1} \sum_{M_i, M_f} \left| \langle \Psi_f | T | \Psi_i \rangle \right|^2
\end{equation}
The summation is over the transition matrix elements represented by M.
More rigorous treatment of MCDHF is provided in \cite{grant2007relativistic}
We utilized this theory for the calculation of atomic orbitals; we then used the RDW method with these orbitals for the construction of the transition matrix and calculation of EIE.

\subsection{RDW-method}
According to RDW theory, the Transition matrix is given by \cite{GUPTA2020106992}

 
 \begin{multline}
T_{l \to u}^{\text{RDW}}(J_u M_u, \vec{k}_u \mu_u ;  J_l M_l, \vec{k}_l \mu_l, \theta) = \\
\langle \phi_{u}^{rel}(1,2,\dots ,N) F_{u,\mu_u}^{DW-}(\vec{K_u},N+1)|V-U_f| \\
\times A \phi_{l}^{rel}(1,2,\dots,N)F_{l,u_l}^{DW+}(\vec{k_l},N+1) \rangle
\end{multline}

here $\phi$ represent bound state wave functions ( $l$ and $u$ denotes lower and upper state) , $J$ and $M$ are total angular momentum and corresponding magnetic component.
 $F_{l(u),\mu_{l(u)}}^{DW+} $ is a relativistic distorted wave function for incoming (outgoing) electron having wave vector $\vec{k_{l(u)}}$ having magnetic components $\mu_{l(u)}$, the angle between the incident and scattered electron is given by $\theta$
 and $-(+)$ denotes incoming(outgoing) electron boundary condition. once we have the T-matrix, the EIE cross-section can be calculated as 
\begin{multline}
\sigma_{lu}^{ex}= \sum_{M_u}\sigma_{M_u} = 
\sum_{M_u} \frac{2\pi^2}{(2J_l +1)}\frac{k_u}{k_l}\sum_{\mu_l\mu_u M_l} \\
\times \int \big|T_{l\to u}^{\text{RDW}}(J_u,M_u,\vec{k_u},\mu_u;J_l,M_l,\vec{k_l},\mu_l)\big|^2 d\Omega
\end{multline}

The integration is over the scattered direction of electrons, and summation is performed on final magnetic states, whereas the average is done based on initial magnetic states.  

Now the Rate coefficient can be calculated with \cite{GUPTA2020106992}
\begin{equation}
R_{lu}^{ex}=\sqrt{2}\int_{E_{lu}}^{\infty}\sigma_{lu}^{ex}\sqrt{E}f(E)dE
\end{equation}
where we integrate by taking the lower limit as the excitation threshold energy of transition. f(E) is the electron energy distribution function. Next, we discuss the CR- modeling. 

\subsection{Collisional-Radiative model}
Collisional-Radiative model \cite{JOHNSON2019100579,HARTGERS2001199} involves solving the kinetics equation for a population of considered level when the time scale of collisional and radiative processes involved are smaller than other effects, then it can be assumed that the density change is mainly because of these processes. we considered the following processes for CR-model calculation,
\subsubsection{Electron impact excitation and de-excitation}
The collision between an atom (A) and electron (e) may result in the excitation of atom; the reaction can be represented as 
\begin{equation}
    A_i + e + Q \longleftrightarrow A_f + e 
\end{equation}
here, Q is the energy difference between the two states marked as $i,f$. 
The probability of such excitation depends on the incident electron energy. the reaction coefficient is given by \cite{HARTGERS2001199}
\begin{equation}
    K(p,q) = \int_{E_{pq}}^{\infty}\sigma_{pq}(E)f(E)v_e(E)dE
\end{equation}
where $v_e(E)=\sqrt{2E/m_e}$ ($m_e$ being electronic mass, $E$ is energy) is electron's velocity, $\sigma_{qp}(E)$ is cross-section, $f(E)$ represent electron energy distribution function. 
For the reverse process, we utilize the detail balancing principle to calculate the reaction coefficient. 
we have 
\begin{equation}
    K(q,p)= \frac{g(p)}{g(q)}K(p,q)exp(\frac{E_{pq}}{kT_e})
\end{equation}
here, $g$ is the statistical weight of the corresponding level,k is the Boltzmann constant, and $T_e$ is the electron temperature. One can also show that 
\begin{equation}
    \sigma_{qp}(E)= \frac{g(p)}{g(q)}\frac{E+E_{pq}}{E}\sigma_{pq}(E+E_{pq})
\end{equation}. 
\subsubsection{Electron impact ionization and three-body recombination}
The process can be represented as 
\begin{equation}
    A_p + e + (Q_{p+}) \longleftrightarrow X_{+}+ e + e 
\end{equation}
for the above process, the electron should have a sufficient amount of energy so that ionization is possible; the reverse process is called three-body recombination. 
Under the equilibrium condition, one has 
\begin{equation}
    \frac{n(p)}{g_p}= \frac{n_e}{g_e}\frac{n_+}{g_+}(\frac{h^2}{2\pi m_ekT_e})^{3/2}exp(\frac{E_{p+}}{kT_e})
\end{equation}

\subsubsection{Spontaneous radiative decay}
The last process we considered for our calculation is spontaneous radiative decay 
\begin{equation}
    A_p \longleftrightarrow A_q + hv_{pq}
\end{equation}

By incorporating these three processes, we write the kinetic equation 
as \cite{kr}

\begin{multline}
\sum_{l; l\neq u}K_{lu}^{ex}(T_e)n_ln_e 
+\sum_{l>u}R_{lu}{n_l}+ n_en_+n_ek_{+u}(T_e) \\ 
- \sum_{l; l\neq u}K_{ul}^{de-ex}(T_e)n_un_e\\ 
-\sum_{l<u}R_{ul}n_u-n_un_ek_{u+}(T_e) = 0 
\end{multline}

\vspace{10pt}
Here $+(-)$ represent processes that populate(depopulate) channels. $R$ is radiative transition rates and $K$ denotes rate coefficients , $n$ being the electron density.

\section{Results and discussion }
\begin{table*}[h!]
\centering
    \caption{Excitation energy (ev) for Na-like Krypton.}
    \label{kr energy}
    \begin{tabular}{|c|c|c|c|c||c|c|c|c|c|}
        \hline
        Configuration & $2J$ & Calculated &  NIST & Energy\cite{kr} & Configuration & $2J$ & Calculated &  NIST & Energy\cite{kr} \\
        \hline
        $3p^1$   & 1   & 56.458   & 56.3400  & 56.2050   & $4p^1$   & 1   & 579.8104   & 580.180  & 579.760  \\
        $3p^1$   & 3   & 69.369   & 69.2670   & 69.1252  & $4p^1$   & 3   & 584.9677   & 585.280  & 584.912  \\
        $3d^1$   & 3   & 144.576  & 144.340  & 144.431  & $4d^1$   & 3   & 613.0113   & 613.390  & 612.904  \\
        $3d^1$   & 5   & 146.997  & 146.796  & 146.865  & $4d^1$   & 5   & 614.0897   & 614.460  & 613.977  \\
        $4s^1$   & 1   & 556.807  & 557.150  & 556.622  & $5s^1$   & 1   & 800.2809   & 800.830  & 800.197  \\
        $5p^1$   & 1   & 811.778  & 812.340  & 811.722  & $5d^1$   & 3   & 827.9776   & 828.550  & 827.936  \\
        $5p^1$   & 3   & 814.346  & 814.910  & 814.286  & $5d^1$   & 5   & 828.5371   & 829.110  & 828.493  \\
        $6s^1$   & 1   & 928.297  & 928.910  & 928.199  & $6p^1$   & 1   & 934.8806   & 935.460  & 934.802  \\
        $6p^1$   & 3   & 936.338  & 936.930  & 936.256  & $6d^1$   & 3   & 944.0083   & 944.580  & 943.933  \\
        $6d^1$   & 5   & 944.332  & 944.910  & 944.256  & $7s^1$   & 1   & 1003.9016  & -        & 1003.780 \\
        $7p^1$   & 1   & 1007.998 & -        & 1007.890 & $7d^1$   & 3   & 1013.6493  & -        & 1013.540 \\
        $7p^1$   & 3   & 1008.904 & -        & 1008.800 & $7d^1$   & 5   & 1013.8539  & -        & 1013.740 \\
        \hline
    \end{tabular}
\end{table*}

\begin{figure*}[h!]
  \centering
  \includegraphics[scale=0.60]{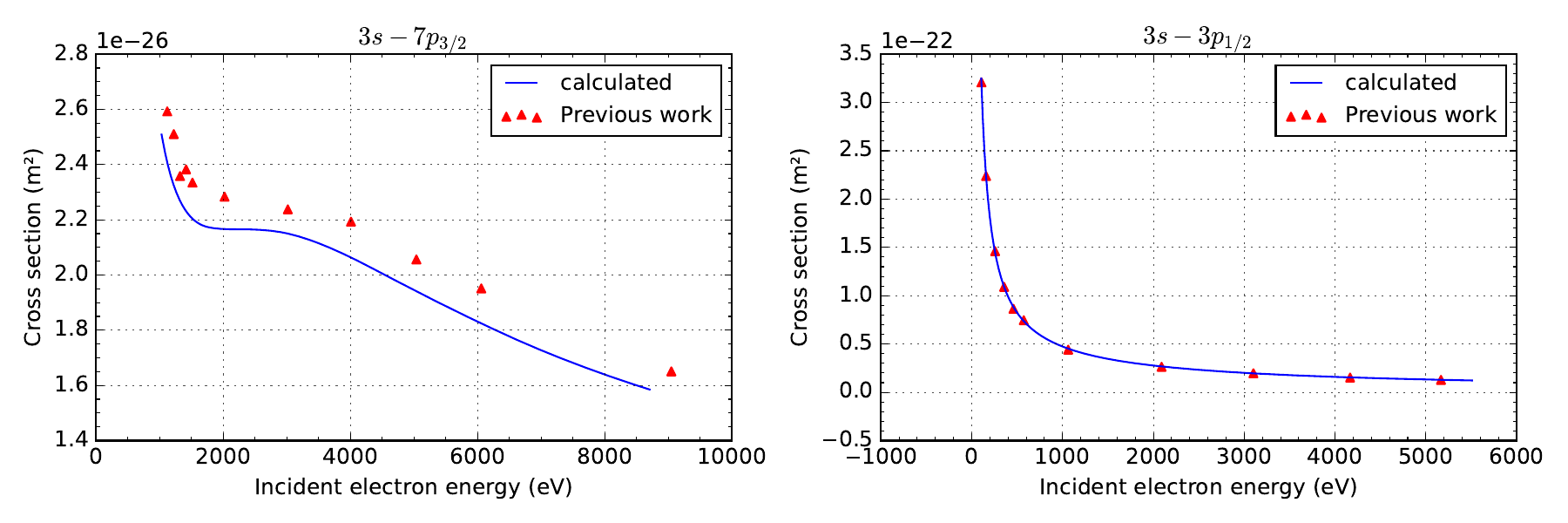}   
  \caption{EIE cross-section comparison for Na-like Kr with previous work \cite{kr}.}
 \label{Kr EIE}
 \end{figure*}
  \begin{figure*}[h!]
  \centering
  \includegraphics[scale=0.60]{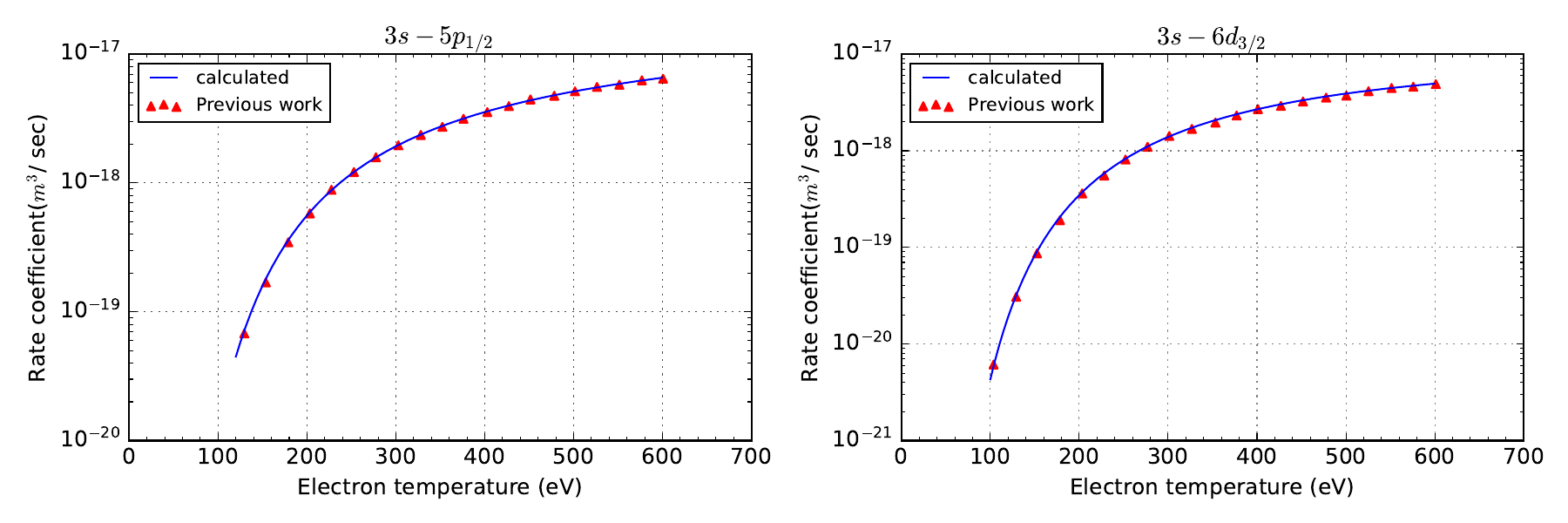}   
  \caption{Comparison of calculated rate coefficient with \cite{kr} for Na-like Kr. }
 \label{Kr rate}
 \end{figure*}

\subsection{$Kr^{25+}$}
  For Na-like ion electronic configuration is $1s^22s^22p^63s$ the excitation included are $3p,3d,nl$ $ (4\leq n \leq 7, 0 \leq l \leq 2)$.
 The calculation for fine structure levels was performed using the MCDHF theory. Finally, 24 fine structure levels were considered. The results for excitation energy are provided in Table\ref{kr energy}; in the table, we compare our result with NIST\cite{nist_atomic_spectra} (National Institute of Standard and Technology) and previous work \cite{kr},

 \begin{figure*}[h!]
  \centering
  \includegraphics[width=15.5cm, height=6.8cm]{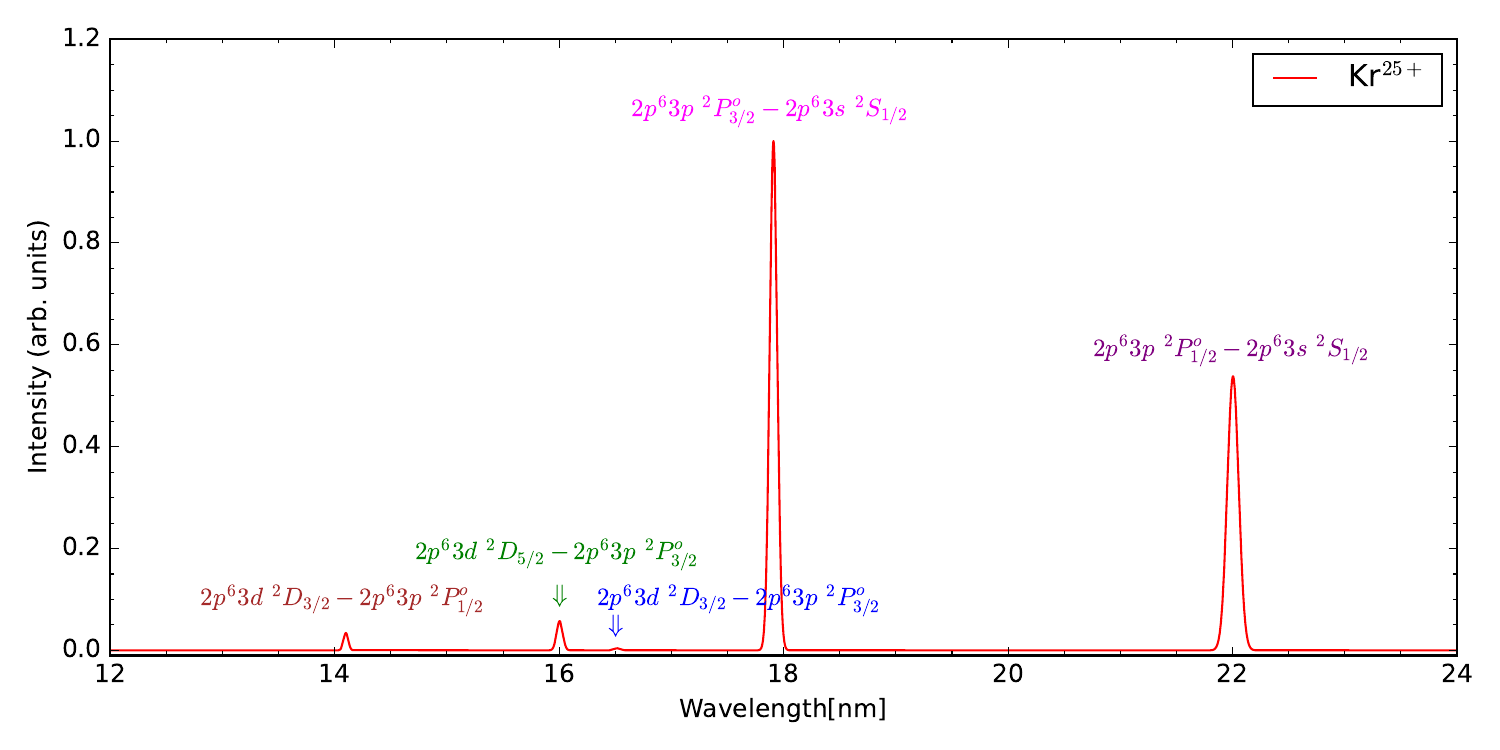}   
  \caption{Emission spectra of Kr$^{25+}$.}
 \label{fig7}
 \end{figure*}
  \begin{figure*}[h!]
  \centering
 \includegraphics[width=15.5cm, height=6.8cm]{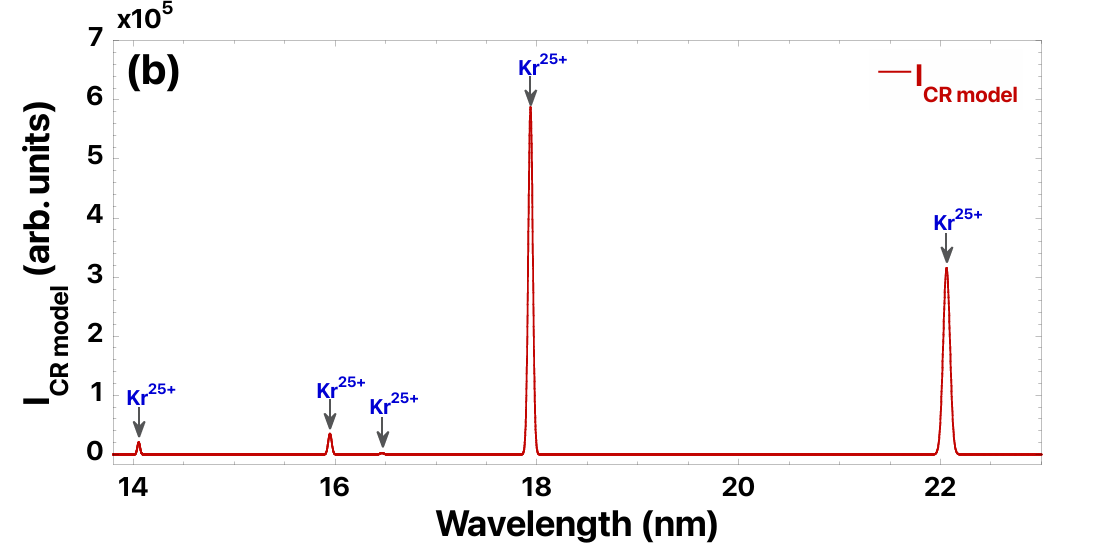} 
  \caption{Emission spectra of Kr$^{25+}$ taken from \cite{kr}.}
 \label{fig8}
 \end{figure*}

 The excitation energy for considered fine structure levels is in agreement with both NIST and \cite{kr}, which has mean percentage difference and variance of NIST(0.015,0.008), \cite{kr}(0.050,0.011).

\begin{figure}[h!]
  \flushleft  
  \includegraphics[width=8.5cm, height=5.6cm]{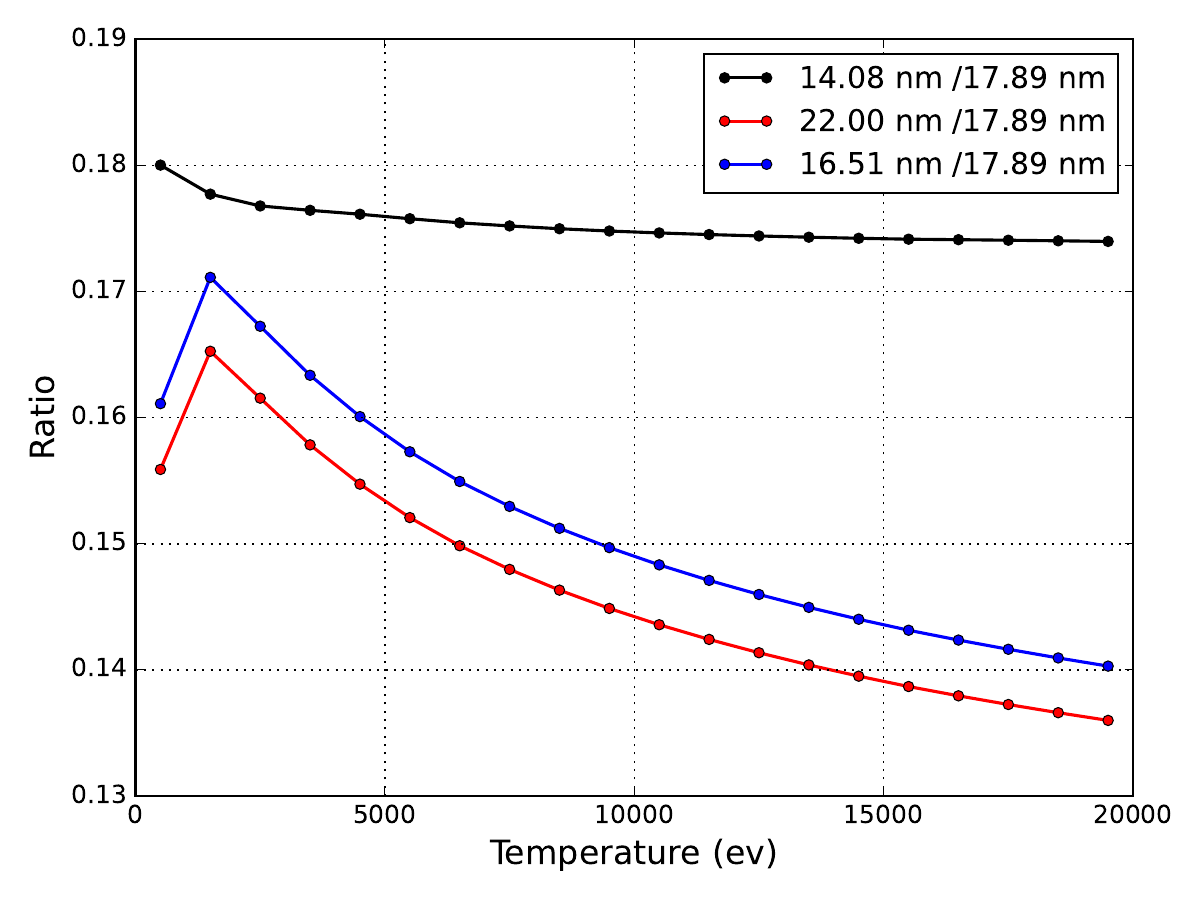} 
  \caption[Variation of line ratios with temperature for Na-like Kr.]{\raggedright Variation of line ratios with temperature for Na-like Kr.}
  \label{line_kr}
\end{figure}

 For the calculation of EIE, we used the RDW method, and the results are shown in Fig\ref{Kr EIE} for $3s-7p_{3/2}$ and $3s-3p_{1/2}$.
These two transitions are shown to have a picture of two extremes. The computed cross-section are also in agreement. Fig\ref{Kr rate} provide the rate coefficient for $3s-5p_{1/2}$ and $3s-6d_{3/2}$.
\\
Rate Coefficients are in remarkable agreement, which further strengthens our calculation. 
Next, we move to the discussion of the CR model; for calculation in the CR model for Kr, we considered the three processes as described in theory; besides the excitation levels already mentioned, we included the following ionization state: $1s^2 2s^2 2p^6$, $1s^ 2s^2 2p^5 3s$, $1s^2 2s^1 2p^6 3s$, $1s^1 2s^2 2p^6 3s$ for calculation. we solve for the population of levels(as described in the theory of CR-model), once we have a population of levels, the intensity of emission lines for transition $ u \rightarrow l $ can be calculated as 
$I_{ul} = E_{lu}R_{ul}n_u$, In the whole CR-model calculation we considered parameters(density of ion , electron temperature) as mentioned in \cite{kr}.  The computed intensity profile is shown in fig \ref{fig7}, these lines correspond to \cite{kr} $2p^63p( \prescript{2}{}{P_{1/2}^{o}})-2p^63s(\prescript{2}{}{S_{1/2}})$, $2p^63p( \prescript{2}{}{P_{3/2}^{o}})-2p^63s(\prescript{2}{}{S_{1/2}})$, $2p^63d( \prescript{2}{}{D_{3/2}})-2p^63p(\prescript{2}{}{P_{3/2}^{o}})$, $2p^63d( \prescript{2}{}{D_{5/2}})-2p^63p(\prescript{2}{}{P_{3/2}^{o}})$, and $2p^63d( \prescript{2}{}{D_{3/2}})-2p^63p(\prescript{2}{}{P_{1/2}^{o}})$ which are at 22.00,17.89,16.51,15.99, and 14.08 nm respectively shown in fig\ref{fig8}. Our calculation is consistent with \cite{kr}. Next we used  this CR model to compute the variation of line ratio for Na-like Kr with temperature, the results are shown in \ref{line_kr}. 

\begin{table*}[h!]
    \centering
    \caption{Excitation energy(ev) for Na-like Xenon.}
    \label{Xe energy}
    \begin{tabular}{|c|c|c|c|c|c|c|c|c|c|c|c|c|c|c|}
        \hline
        Configuration & 2J & Calculated &  NIST &Configuration & 2J & Calculated &  NIST & Configuration & 2J & Calculated &  NIST \\
        \hline
        $3p^1$   & 1   & 100.20 & 100.00  &
        $3p^1$   & 3   &  186.29  & 186.13 &
        $3d^1$ & 3 & 313.14 & 312.89\\
        $3d^1$ & 5 & 332.36& 332.20 &
        $4s^1$ & 1 & 1519.6 & 1520.4&
        $4p^1$ & 1 &1561.0& 1561.7\\ 
        $4p^1$ & 3 &1596.1 & 1596.9&
        $4d^1$ & 3 & 1644.0& 1644.0&
        $4d^1$ & 5 & 1652.4 & 1652.8\\
        $5s^1$  & 1 & 2194.6 & 2195.5&
         $5p^1$ & 1 &2215.5 &2216.4&
       $5p^1$ & 3 &2233.0&2234.0 \\
        $5d^1$ & 3 &2256.5 &2257.4 &
        $5d^1$ & 5 &2260.8 &2261.8 &
        $6s^1$ & 1 &2552.2 &2552.6 \\
        $6p^1$ & 1 &2564.3 &2564.5&
        $6p^1$ & 3 &2574.4 &2574.6 &
        $6d^1$ & 3 &2587.6 &2587.9 \\
        $6d^1$ & 5 &2590.1 &2590.4 &
        $7s^1$ & 1 &2764.5 & - &
        $7p^1$ & 1 &2772.2 & - \\
        $7p^1$ & 3 &2778.3 & - &
        $7d^1$ & 3 &2786.5 & - &
        $7d^1$ & 5 &2788.1 & - \\
         \hline
    \end{tabular}
\end{table*}

 \begin{figure*}[h!]
  \centering
  \includegraphics[scale=0.60]{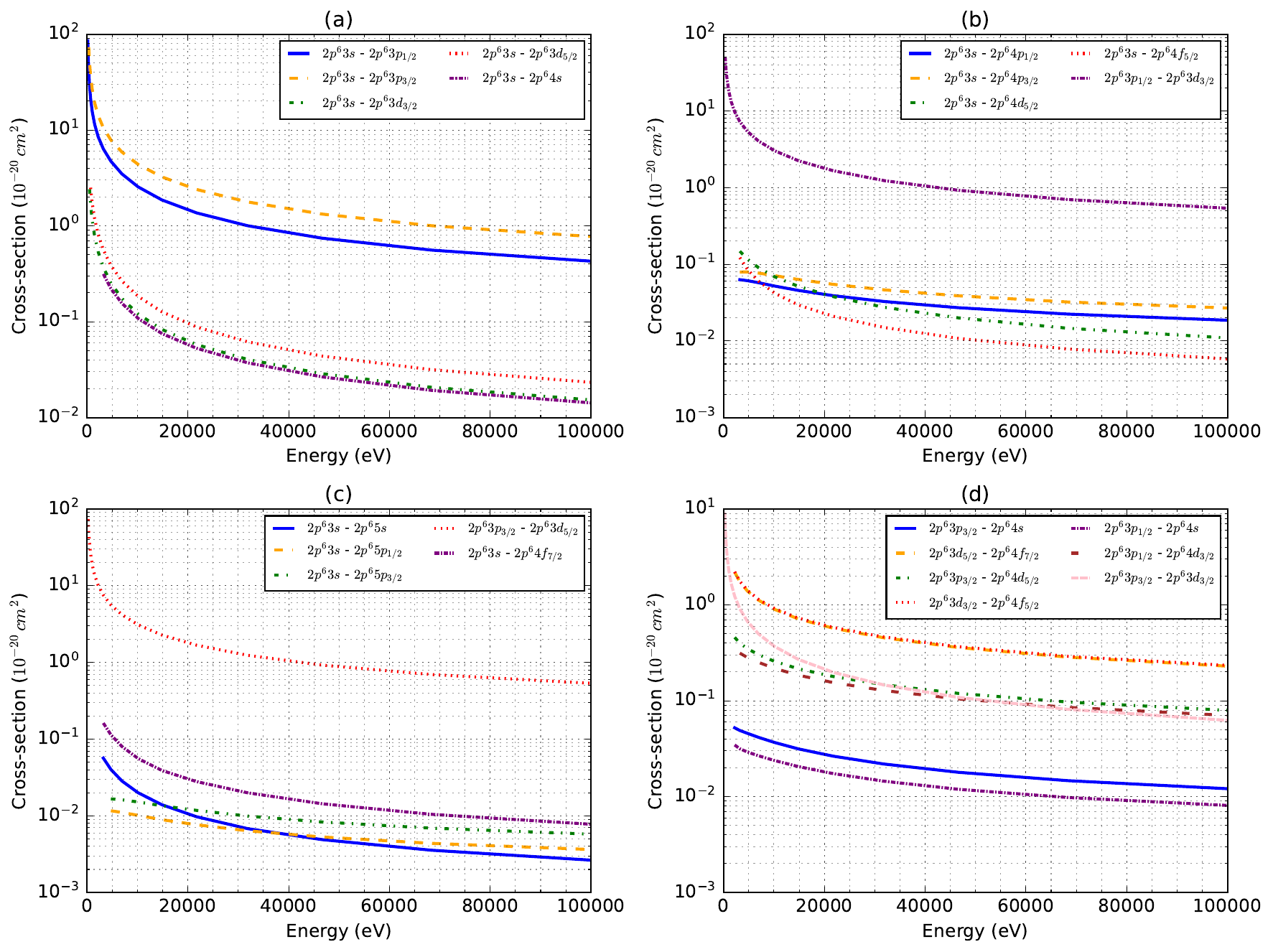}   
  \caption{EIE cross-section variations with electron energy.}
 \label{Xe cross}
 \end{figure*}
For intensity at 14 nm and 16.51 nm the ratios increases initially and then decreases, contrast to that for 22.00 the line ratio decreases continuously.

\subsection{$Xe^{43+}$}
In the previous section we have validated our results by comparing the computed values with the previous work. In this section we provide our result for Na-like Xe. The levels considered for calculation were same as in our previous work \cite{article}. We provide the EIE cross-section, line emission intensity, and variation of line ratios with temperature.

 \begin{figure*}[h!]
  \centering
  \includegraphics[width=15.5cm, height=8cm]{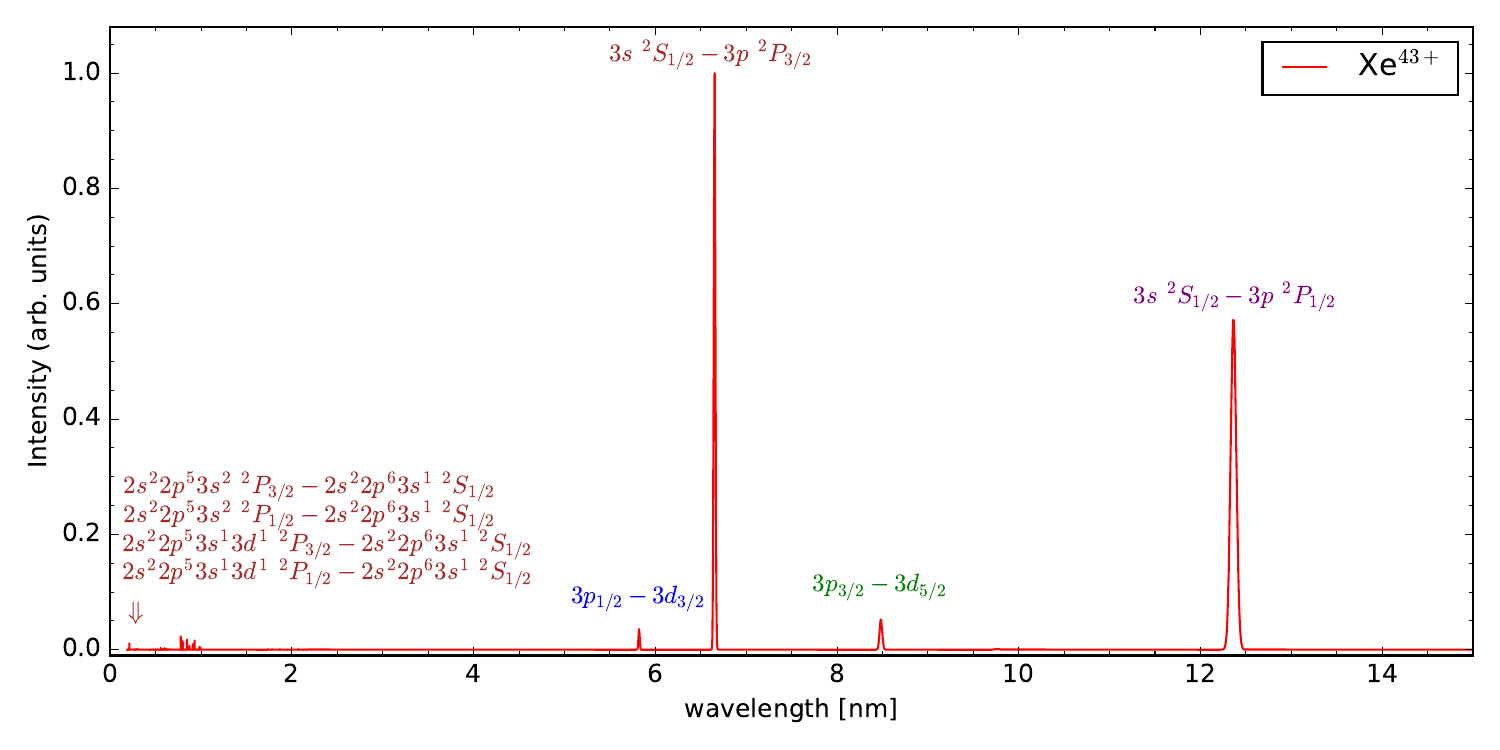}   
  \caption{Emission spectra for Xe$^{43+}$.}
 \label{xe inten}
 \end{figure*}
 \begin{figure*}[h!]
  \centering
  \includegraphics[width=16cm, height=6.8cm]{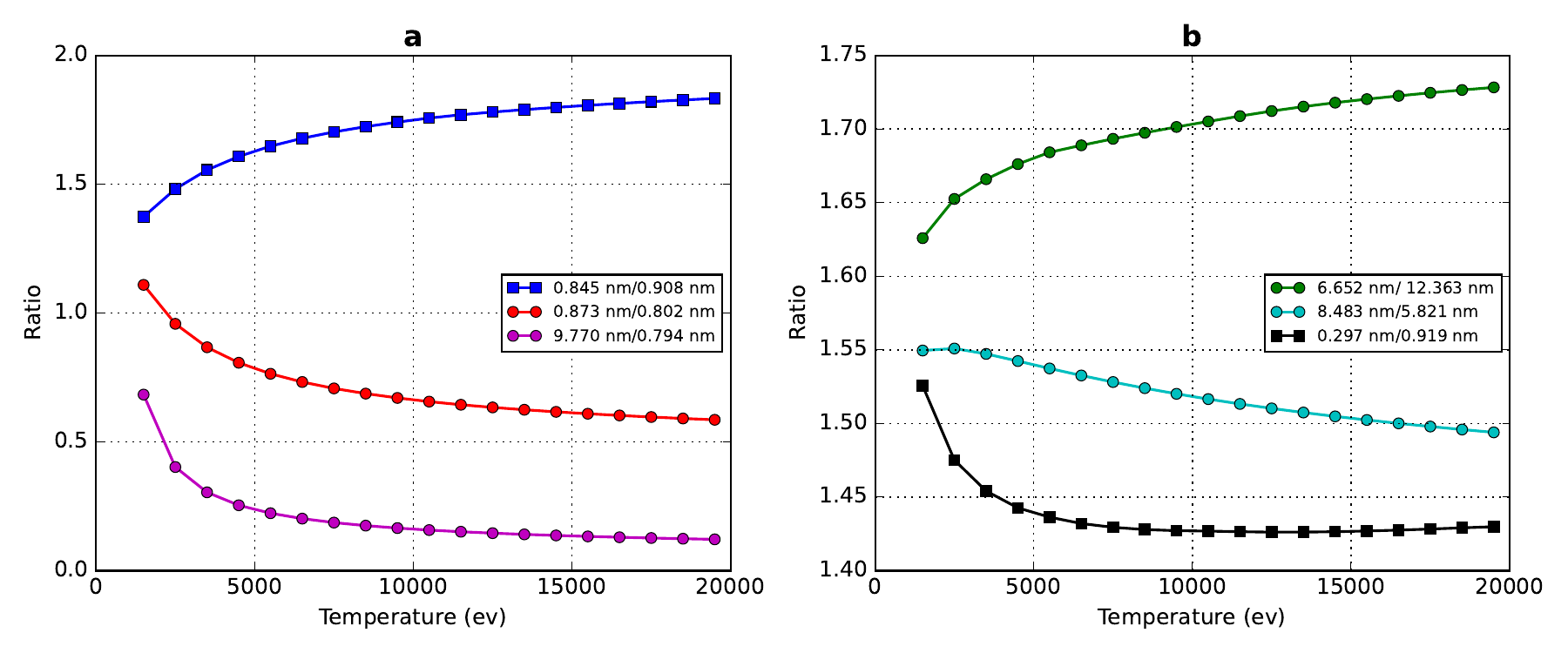}   
  \caption{Variation of line ratios with temperature for Na-like Xe.}
 \label{line ratio}
 \end{figure*}

Table \ref{Xe energy} compares fine structure levels with NIST data. We obtained a mean percentage difference of 0.0011 and a variance of 0.003, Hence our excitation energies are in accordance with NIST.  Fig\ref{Xe cross} provides the EIE cross-section for various transitions, the corresponding data is also provided in form of table in appendix \ref{Xe_cross_table}.

These transitions were present in the intensity spectra of our CR-model.
The result of the CR model is shown as an intensity profile in Fig \ref{xe inten}. We have two sets of lines. One is in the Nanometer range, and another is in the angstrom region. In the nanometer region, the prediction of our model agrees with previous experiments and predictions  \cite{PhysRevA.68.042501} 8.48, 5.82,6.61, 6.66 (we have one intensity corresponding for 6.61 and 6.66 due to the Gaussian convolution), \cite{Osin_2012}12.39.

In the Angstrom range, we also see some intensities that correspond to \cite{https://doi.org/10.1002/xrs.850} 2.96 2.74; along these intensities, we have a few more in the Angstrom region. 
The calculation of the CR model was performed for various electron temperatures, and the variations of line ratio are provided in Fig \ref{line ratio}.

\section{Conclusion}
In this work, we utilized the MCDHF and RDW methods for our calculation of atomic parameters and EIE calculation for Na-like ion Kr and Xe.
To check the reliability of our calculation, we compare the results with available data for excitation energy of fine structure, EIE cross-section, and rate coefficients for Na-like Kr. We obtained a good agreement for our calculation. Next, we provided results for Na-like Xenon. The EIE-cross, intensity spectra, and line ratios, given for Na-like Xe, can provide insight into understanding experimental astronomical plasma. The intensity spectra that we obtained from our CR model for Na-like Xenon are in agreement with previous theoretical and experimental work. Spectroscopic studies involving emission lines can be used to diagnose plasma. Moreover, these results can provide a deeper understanding of atomic and plasma theory.

\appendices
\onecolumn
\begin{landscape}
\section{EIE Corss-section( $10^{-20} cm^2$) for various transitions correspoinding to energy(ev). }
 \label{Xe_cross_table}
\begin{longtable}{|c|c|c|c|c|c|c|c|c|c|c|c|c|}
\hline
 &  Energy &   $3s$ - $3p_{1/2}$ &   $3s$ - $3p_{3/2}$ &   $3s$ - $3d_{3/2}$ &   $3s$ -$3d_{5/2}$ &  $3s$ - $4s$ &   $3s$ - $4p_{1/2}$ &   $3s$ - $4p_{3/2}$ &   $3s$ - $4d_{5/2}$ &   $3s$ - $4f_{5/2}$ &   $3s$ - $4f_{7/2}$ &   $3s$ - $5s$ \\
\hline
   0 &    104.24 &                       122.81    &                         0       &                        0        &                        0        &                  0        &                        0        &                        0        &                        0        &                       0         &                       0         &                 0         \\
   1 &    105.03 &                       121.92    &                         0       &                        0        &                        0        &                  0        &                        0        &                        0        &                        0        &                       0         &                       0         &                 0         \\
   2 &    105.84 &                       121.03    &                         0       &                        0        &                        0        &                  0        &                        0        &                        0        &                        0        &                       0         &                       0         &                 0         \\
   3 &    106.65 &                       120.15    &                         0       &                        0        &                        0        &                  0        &                        0        &                        0        &                        0        &                       0         &                       0         &                 0         \\
\hline
\end{longtable}

\begin{longtable}{|c|c|c|c|c|c|c|c|c|c|c|c|c|}
\hline
     &  Energy &  $3s$-$5p_{1/2}$ &   $3s$-$5p_{3/2}$ &  $3p_{3/2}$-$3d_{5/2}$ &   $3p_{1/2}$-$3d_{3/2}$ &  $3p_{3/2}$-$4s$ &   $3d_{5/2}$-$4f_{7/2}$ &   $3p_{3/2}$-$4d_{5/2}$ &   $3d_{3/2}$-$4f_{5/2}$ &   $3p_{1/2}$-$4s$ &  $3p_{1/2}$-$4d_{3/2}$ &  $3p_{3/2}$-$3d_{3/2}$ \\
\hline
   0 &    104.24 &                       0         &                       0         &                               0       &                               0       &                        0        &                               0       &                              0        &                               0       &                       0         &                              0        &                              0        \\
   1 &    105.03 &                       0         &                       0         &                               0       &                               0       &                        0        &                               0       &                              0        &                               0       &                       0         &                              0        &                              0        \\
   2 &    105.84 &                       0         &                       0         &                               0       &                               0       &                        0        &                               0       &                              0        &                               0       &                       0         &                              0        &                              0        \\
\hline
\end{longtable}
\end{landscape}
\twocolumn
\bibliographystyle{IEEEtran} 
\bibliography{Mybib.bib}
\end{document}